\documentclass[aps,
prd,
reprint,
amsmath,
amssymb,
amsthm, 
bm,
]{revtex4-2}

\usepackage{graphicx}
\usepackage{dcolumn}
\usepackage{bm}
\usepackage{multirow} 
\usepackage{empheq}
\usepackage{xcolor}
\usepackage{soul}
\usepackage{hyperref}
\usepackage{amsmath, amsthm, amssymb, bm} 
\usepackage{placeins} 
\renewcommand\hl[1]{#1}

\begin{document}
	
	\title{A statistical inference approach to time-delay interferometry for gravitational-wave detection}
	
	
	\author{Quentin Baghi}
	\email[]{quentin.s.baghi@nasa.gov}
	\author{James Ira Thorpe}
	\author{Jacob Slutsky}
	\author{John Baker}
	
	\affiliation{Goddard Space Flight Center \\
		Mail Code 663\\
		8800 Greenbelt Rd, Greenbelt, 
		Maryland 20771, USA}
	
	
	\date{\today}
	
	\begin{abstract}
		The future space-based gravitational wave observatory LISA will consist of a constellation of three spacecraft in a triangular constellation, connected by laser interferometers with 2.5 million-kilometer arms. Among other challenges, the success of the mission strongly depends on the quality of the cancellation of laser frequency noise, whose power lies eight orders of magnitude above the gravitational signal. The standard technique to perform noise removal is time-delay interferometry (TDI). TDI constructs linear combinations of delayed phasemeter measurements tailored to cancel laser noise terms. Previous work has demonstrated the relationship between TDI and principal component analysis (PCA). We build on this idea to develop an extension of TDI based on a model likelihood that directly depends on the phasemeter measurements. Assuming stationary Gaussian noise, we decompose the measurement covariance using PCA in the frequency domain. We obtain a comprehensive and compact framework that we call PCI for ``principal component interferometry," and show that it provides an optimal description of the LISA data analysis problem.
	\end{abstract}
	
	
	\maketitle
	

	\section{Introduction} \label{sec:intro}
	Only five years since the historic first detection of an astrophysical signal, the field of gravitational-wave astronomy continues to rapidly advance through upgrades to ground-based facilities~\cite{Aasi2013}, plans for ambitious follow-on terrestrial detectors~\cite{Abbott2017a}, and a steady increase in sensitivity of pulsar timing arrays~\cite{Dahal2020}. Adding to this in the 2030s will be the space mission known as the Laser Interferometer Space Antenna (LISA)~\citep{Amaro-Seoane2017, KarstenDanzmann2017}. LISA will open the source-rich frequency window between 0.1~mHz and 1~Hz, enabling the detection of thousands of gravitational-wave emitting systems of varied origin and at distances ranging from our galactic neighborhood (kpc) to cosmological redshift~($z=15$ and beyond). The types of sources will include from compact binary stars, the capture of stellar-remnant black holes by massive black holes, to mergers of (super)massive black hole binaries. LISA will form a triangular constellation of 3 satellites with trailing Earth on heliocentric orbits. Each satellite will house inertial test-masses whose trajectories will be monitored through a network of interferometric laser links, connecting the spacecrafts separated from each other by 2.5 million kilometers. Incoming gravitational waves (GW) will affect the space-time across the constellation, introducing a characteristic shift in the light travel time along the six one-way optical links.
	
	Due to the large distance between the inertial references, LISA needs \hl{multiple interferometric measurements} to operate optimally~\citep{Otto2012}. In this setup, the phasemeter measurements are sensitive to noise fluctuations of the laser frequencies at a level of $10^{-13} \rm \, {Hz}^{-1/2}$. The typical metric perturbations induced by the target GW sources being about~$10^{-21}$, the laser noise dominates them by eight orders of magnitude. Recovery of the GW signal is possible because the correlations between the individual link signals differ for GWs and laser noise. Laser frequency noise is almost entirely canceled by a post-processing technique called time-delay interferometry (TDI)~\citep{Armstrong1999}, which constructs linear combinations of delayed phasemeter measurements tailored to cancel the laser frequency noise up to ranging errors. Some TDI variables can be interpreted physically by synthetically retracing the path of light rays traveling in a classical Michelson interferometer. The feasibility of TDI has been studied extensively over the past decade~(see, e.g., Ref.~\cite{Tinto2014} for an overview), including the implementation of interpolation filters needed to produce the laser-free data streams~\citep{Shaddock}. Further analyses have recently been tackled, revealing expected TDI noise artifacts, such as flexing-filtering~\citep{Bayle2019} and clock jitters effects~\citep{Hartwig2020}.
	
	Different generations of TDI variables achieving different accuracy levels can be formed depending on assumptions about the spacecraft motion. TDI generations 1.0, 1.5, and 2.0, respectively, assume a rigid and static constellation, a rigid and rotating configuration, and a flexing, rotating configuration where the armlengths are linearly varying in time. Recent work also identified new TDI combinations by using the explicit dependence of arm delays on the satellites' velocities and accelerations~\cite{Muratore2020}.
	
	In addition to its geometric interpretation, TDI can be viewed algebraically as the group of solutions of an equation involving six-tuples polynomials in the delay operators~\citep{Nayak2004}, which encodes the cancellation of noise. Beyond laser noise cancellation, the analysis of LISA's measurement can also be interpreted as an inference problem. We observe the same physical effect (the perturbation of the metric due to incoming gravitational waves) through different sensors (interferometric arms) affected by fundamental errors (the laser noise and other stochastic noises). Everything works as if we recorded the same sound using different microphones.
	The goal of the analysis is generally to estimate parameters of astrophysical sources emitting GWs. Thus, we can infer them by directly writing the likelihood to observe the phasemeter measurements given a prescribed model. The model describes how we expect the GW signal to appear in the various measurements and the relationships between the noises present in each data stream. In Ref.~\cite{Vallisneri2020}, Vallisneri et al. recently formalized this idea in the time domain, by marginalizing the likelihood with respect to laser noises. In this study, we develop a similar approach in the frequency domain, using an eigen-decomposition of the covariance.
	
	Typically, the likelihood depends on the inverse correlation matrix of all measured variables. Applying its inverse to the model residuals yields the weighed, uncorrelated squared errors involved in any optimal parameter estimation scheme. We refer to the process of generating orthogonal variables as principal component analysis (PCA). In a pioneering work~\cite{Romano2006}, Romano and Woan show that we can derive the TDI combinations from the eigenvectors of the single-link covariance matrix through a demonstration based on a simplified time-domain analysis assuming white noise and short time series. \hl{Leighton further analyzes the noise covariance matrix structure and its principal components in Ref.}~\cite{Leighton2016}, \hl{extending it to the frequency domain. The aim of the present work is to make this idea readily usable for gravitational-wave data analysis and demonstrate it on an inference problem.} For this purpose, we build an analytic formalism implementing PCA, based on a matrix formulation in the frequency domain. We refer to this approach as ``Principal Component Interferometry" or PCI for short.
	
	PCI implements the generalized analog of the orthogonal TDI channels A, E T~\citep{Prince}. While A, E, T's original construction relies on assumptions such as equal arms and uncorrelated acceleration noises, PCI yields the optimal variance in a unified, data-driven formalism. We first decompose the data on the eigenvector basis of the laser-noise covariance matrix. We then estimate the frequency-domain covariance matrix (or spectral matrix) of eigenstreams, assuming that the acceleration noise is stationary. The resulting model likelihood allows us to fit for noise parameters encoding all types of noise correlations, which we account for by construction in the inference process.
	
	In Sect.~\ref{sec:formalism}, we present the algebraic formalism used to develop PCI. In Sect.~\ref{sec:pca}, we derive the principal components of the covariance in the simple case of rigid armlengths. In Sect.~\ref{sec:simulations}, we demonstrate PCI's performance through simple numerical simulations, with an example of an application where we estimate conjointly the laser light propagation delays (ranging), the noise covariance elements, and the parameters of a compact galactic binary source. We conclude and discuss the further generalization to more complex cases in Sect.~\ref{sec:discussion}.

	\section{\label{sec:formalism}Modeling single-link measurements using matrix operators}
	
	\subsection{Derivation of the likelihood}
	In this section, we present the conventions and the formalism that we adopt throughout the study. Using conventions in Fig.~1 of Ref.~\citep{Bayle2019}, we consider a spacecraft labeled $i$, and the signal $s_{i}$ obtained by comparing the laser light coming from the distant spacecraft $i+1$ to the local oscillator of optical bench $i$. The measurement $s_{i}$ is where gravitational waves imprint their presence and is called the science interferometer signal. 
	The science interferometer measurements $s_{i}(t) = \left(\nu_{i}(t) - \nu_{0} \right) / \nu_{0}$ are expressed as the relative deviation of instantaneous frequency with respect to the carrier frequency $\nu_{0}$. For the two optical benches onboard spacecraft $i$, we have, at each time $t$:
	\begin{eqnarray}
		\label{eq:phasemeneter_meas}
		s_{i} &=& h_{i+2} + \mathcal{D}_{i+2} p_{i'+1} - p_{i} + n_{i} ; \nonumber \\
		s_{i'} &=& h_{i+1} + \mathcal{D}_{i'+1} p_{i+2} - p_{i'} + n_{i'},
	\end{eqnarray}
	where $\mathcal{D}_{i}x(t) = x\left(t - c^{-1}L_{i} \right)$ denotes the operator applying the light travel time delay $c^{-1} L_{i}$ along arm $i$, $h_{i}$ is the integrated frequency shift along arm $i$ due to incoming gravitational waves, $p_{i}$ is the frequency noise contribution of the laser in optical bench $i$, and $n_{i}$ gathers all other noises affecting the science measurement on that bench for all $i \in \left\{1, 2, 3\right\}$. We adopt the convention of cyclic indexing where $p_{i}$ actually means $p_{i - 3\left\lfloor (i-1)/3 \right\rfloor}$, and simple indices refer to links and light travel times pointing clockwise, whereas prime indices denote counterclockwise directions. In the following, we assume that the lasers in the two optical benches are identical, so that
	\begin{eqnarray}
		\label{eq:equal_ob_laser_assumption}
		p_{i} & = & p_{i'}  \,\, \forall i \in \left\{1, 2, 3\right\}.
	\end{eqnarray}
	This assumption allows us to simplify the analysis, but can be adopted without loss of generality. Note that although the laser sources are different in reality, they will be compared through reference interferometer measurements.

	The measurements $s_{i}$ in Eq.~(\ref{eq:phasemeneter_meas}) will be sampled at a cadence of $f_{s} = 2$ Hz or more over a finite duration $T$. The resulting time series can be represented by a column vector $\boldsymbol{s}_{i}$ of size $N = f_{s} T$. In this discretized version of the measurement, the delay operators $\mathcal{D}_{i}$ acting on any variable $x(t)$ can be represented by $N \times N$ matrices depending on the arm lengths $L_{i}$. Using consistent notation, all lines in Eq.~(\ref{eq:phasemeneter_meas}) can be re-written as
	\begin{eqnarray}
		\label{eq:observation_equation}
		\boldsymbol{y} = \boldsymbol{h} + \boldsymbol{M} \boldsymbol{p} + \boldsymbol{n},
	\end{eqnarray}
	where $\boldsymbol{y} \in \mathbb{R}^{6N}$ is the column vector stacking the phasemeter measurements for all spacecrafts and all optical benches so that \hl{$y\left( (i-1)N + j\right) = s_{i}\left( j / f_s \right)$ and $y\left((i+2)N + j\right) = s_{i'}\left( j / f_s \right)$ $\forall i \in \left\{1, 2, 3\right\}, \, \forall j \in \left[0,\, N-1 \right]$}.
	Vectors $\boldsymbol{h}$, $\boldsymbol{p}$, and $\boldsymbol{n}$ respectively represent the GW signals, laser noises, and other noises. They have in general the same structure and dimensions as $\boldsymbol{y}$. Note that under the assumption outlined in Eq.~(\ref{eq:equal_ob_laser_assumption}), we keep only the first half of $\boldsymbol{p}$ so that $\boldsymbol{p} \in \mathbb{R}^{3N}$.

	Matrix $\boldsymbol{M}$ encodes the mixing and delaying of laser noises and can be written as a block matrix:
	\begin{eqnarray}
		\label{eq:delay_op_matrix}
		\boldsymbol{M} &=& \begin{pmatrix}
			-\boldsymbol{I}_{N}             & \boldsymbol{D}_{3} & \boldsymbol{0}_{N} \\
			\boldsymbol{0}_{N}             & -\boldsymbol{I}_{N}  & \boldsymbol{D}_{1} \\
			\boldsymbol{D}_{2}   & \boldsymbol{0}_{N}          & -\boldsymbol{I}_{N} \\
			-\boldsymbol{I}_{N}             & \boldsymbol{0}_{N}          & \boldsymbol{D}_{2'}  \\
			\boldsymbol{D}_{3'}   & -\boldsymbol{I}_{N}         &  \boldsymbol{0}_{N}  \\
			\boldsymbol{0}_{N}   & \boldsymbol{D}_{1'}         & -\boldsymbol{I}_{N}  
		\end{pmatrix}.
	\end{eqnarray}
	
	Now that the observation equation is written in matrix form, we can derive the corresponding likelihood.
	Gravitational-wave source parameters $\boldsymbol{\theta}$ are usually extracted using Bayesian inference, which estimates their posterior distribution given the data:
	\begin{eqnarray}
		\label{eq:posterior}
		p\left( \boldsymbol{\theta} | \boldsymbol{y} \right) = \frac{ p\left(\boldsymbol{y}  |  \boldsymbol{\theta} \right)  p\left(\boldsymbol{\theta} \right) }{ p\left(\boldsymbol{y}\right) },
	\end{eqnarray}
	where $p\left(\boldsymbol{y}  |  \boldsymbol{\theta} \right)$ is the model likelihood, $p\left(\boldsymbol{\theta} \right)$ is the prior distribution of the parameters, and $p\left(\boldsymbol{y}\right)$ is the evidence, acting as a normalization.
	
	Assuming a zero-mean Gaussian distribution for all noises, the likelihood follows from Eq.~(\ref{eq:observation_equation}):
	\begin{eqnarray}
		\label{eq:likelihood}
		p\left( \boldsymbol{y}  |  \boldsymbol{\theta} \right) = \frac{\exp\left\{- \frac{1}{2} \left(\boldsymbol{y} -  \boldsymbol{h} \right)^{\dagger}\boldsymbol{\Sigma}^{-1}\left(\boldsymbol{y} -  \boldsymbol{h} \right)\right\}}{\sqrt{(2 \pi)^{6N}  \left| \boldsymbol{\Sigma} \right| }},
	\end{eqnarray}
	where $\dagger$ denotes the Hermitian conjugate, and $\boldsymbol{\Sigma}$ is the $6N \times 6N$ covariance matrix of the observations $\boldsymbol{y}$, whose expression derives from Eq.~(\ref{eq:observation_equation}):
	\begin{eqnarray}
		\label{eq:covariance_matrix}
		\boldsymbol{\Sigma} = \boldsymbol{M} \boldsymbol{\Sigma}_{p} \boldsymbol{M}^{\dagger}+ \boldsymbol{\Sigma}_{n},
	\end{eqnarray}
	where $\boldsymbol{\Sigma}_{p}$ and $\boldsymbol{\Sigma}_{n}$ are respectively the covariance matrices of the laser noises $\boldsymbol{p}$ and of the other noises $\boldsymbol{n}$, assuming no intrinsic correlation between the two. 
	
	In the following, we make the convenient but realistic assumption that all noises are stationary (at least for a relatively short period of time). In that case, their covariance matrices are Toeplitz, and for a sufficiently large $N$, they are approximately diagonalizable in the discrete Fourier basis $\boldsymbol{W}$ which form its eigenvectors and writes $W(k,n) = e^{\frac{2 \pi j nk}{N}}$ with $j = \sqrt{-1}$ being the complex number. We can therefore re-write the covariance matrix in Eq.~(\ref{eq:covariance_matrix}) in the Fourier domain as
	\begin{eqnarray}
		\label{eq:covariance_matrix_fourier}
		\boldsymbol{\tilde{\Sigma}} = \boldsymbol{\tilde{M}} \boldsymbol{S}_{p} \boldsymbol{\tilde{M}}^{\dagger}+ \boldsymbol{S}_{n},
	\end{eqnarray}
	where $\boldsymbol{S}_{p}$ and $\boldsymbol{S}_{n}$ are the covariance matrices of the discrete Fourier-transformed data, that we call spectral matrices. Due to stationarity, $\boldsymbol{S}_{p}$ (respectively $\boldsymbol{S}_{n}$) includes $3 \times 3$ (respectively $6 \times 6$) blocks which are $N \times N$  diagonal matrices, whose diagonal elements are given by the noise cross-spectra. Spectral matrices are Hermitian, with real positive diagonal blocks and complex conjugate off-diagonal blocks. If we restrict the analysis to a specific set of $N_{f}$ frequencies, then each block has size $N_{f} \times N_{f}$. In the following, we look for the principal components of the covariance matrix.\\

	\section{\label{sec:pca}Principal component analysis of single-link measurements}
	
	\subsection{Principle of PCA}
	PCA aims at finding a transformation of the observations that converts them into uncorrelated variables, ordered according to their variance. The process is often used to reduce the dimension of the problem by discarding the highest variance components. 
	Here, we aim at finding a unitary transformation matrix $\boldsymbol{V}$ where the covariance matrix can be diagonalized as
	\begin{eqnarray}
		\label{eq:covariance_diagonalization}
		\boldsymbol{\Sigma} = \boldsymbol{V} \boldsymbol{\Lambda} \boldsymbol{V}^{\dagger},
	\end{eqnarray}
	where $\boldsymbol{\Lambda}$ is a diagonal matrix. 
	Then the log-likelihood can conveniently be re-written as
	\begin{eqnarray}
		\label{eq:likelihood_pca}
		\log p\left( \boldsymbol{y}  |  \boldsymbol{\theta} \right) & = & - \frac{1}{2} \left(\boldsymbol{y} -  \boldsymbol{h} \right)^{\dagger} \boldsymbol{V} \boldsymbol{\Lambda}^{-1} \boldsymbol{V}^{\dagger} \left(\boldsymbol{y} - \boldsymbol{h} \right) \nonumber \\
		&&  - \frac{1}{2} \log \left| \boldsymbol{\Lambda} \right|.
	\end{eqnarray}
	
	However, finding a full decomposition like Eq.~(\ref{eq:covariance_diagonalization}) can be tricky, unless we make a few key assumptions, which we do in the following.

	\subsection{\label{sec:laser_noise_only}PCI for equal noises}
	
	A way to find a decomposition of the form (\ref{eq:covariance_diagonalization}) is to find the eigenvectors of the covariance matrix. To ease their calculation, we make two assumptions.
	
	First, we assume that all delays are constant in time, which implies that the delay operators are commutative, i.e. $\boldsymbol{D}_{1}\boldsymbol{D}_{2} = \boldsymbol{D}_{2} \boldsymbol{D}_{1}$. In this case, the Fourier basis also provides approximate eigenvectors for the delay operators $\boldsymbol{D}_{i}$. Hence, in the Fourier basis, the delay operators are approximately diagonal, and in the limit of large $N$, we can write their elements as:
	\begin{eqnarray}
		\label{eq:asymptotic_delay}
		\tilde{D}_{i | k, l} = e^{-2 \pi j f_{k} c^{-1} L_{i}} \delta_{kl}.
	\end{eqnarray}
	
	Second, we assume for now that all non-laser noises have the same power spectral density (PSD). Thus, all diagonal blocks $\boldsymbol{S}_{n,i}$ of the spectral matrix $\boldsymbol{S}_{n}$ are equal: $\boldsymbol{S}_{n,i} = \boldsymbol{\lambda}_{n} \forall i$.
	
	Based on these assumptions, from the calculation of the characteristic polynomial of matrix $\boldsymbol{\tilde{\Sigma}}$ we find that there are 6 eigenvalues per frequency bin, hence $6N_{f}$ eigenvalues for the full problem. Half of them (which we label $\boldsymbol{\Lambda}_{n}$) are degenerate and equal to the non-laser noise PSD values, so that $\boldsymbol{\Lambda}_{n} = \mathrm{diag}\left(\boldsymbol{\lambda}_{n}, \boldsymbol{\lambda}_{n}, \boldsymbol{\lambda}_{n} \right)$. We can write the $6 N_{f} \times 3N_{f}$ matrix gathering their associated eigenvectors $\boldsymbol{\tilde{V}}_{n}$ analytically as
	\begin{widetext}
		\begin{equation}
			\label{eq:constant_delay_eigenvectors}
			\boldsymbol{\tilde{V}}_{n} = 
			\begin{pmatrix}
				\boldsymbol{\tilde{D}}_{2}^{\dagger}\left(\boldsymbol{\tilde{D}}_{1'}^{\dagger} \boldsymbol{\tilde{D}}_{1}^{\dagger} - \boldsymbol{I}\right)  & \boldsymbol{\tilde{D}}_{3'}^{\dagger} - \boldsymbol{\tilde{D}}_{1}^{\dagger}\boldsymbol{\tilde{D}}_{2}^{\dagger}                       & \boldsymbol{\tilde{D}}_{2'}^{\dagger} \boldsymbol{\tilde{D}}_{2}^{\dagger} - \boldsymbol{I} \\
				\boldsymbol{\tilde{D}}_{1'}^{\dagger} - \boldsymbol{\tilde{D}}_{2}^{\dagger}\boldsymbol{\tilde{D}}_{3}^{\dagger}                        & \boldsymbol{\tilde{D}}_{3'}^{\dagger}\boldsymbol{\tilde{D}}_{3}^{\dagger} - \boldsymbol{I}                                           & \boldsymbol{\tilde{D}}_{3}^{\dagger} \left( \boldsymbol{\tilde{D}}_{2'}^{\dagger}\boldsymbol{\tilde{D}}_{2}^{\dagger} - \boldsymbol{I}\right) \\
				\boldsymbol{\tilde{D}}_{1'}^{\dagger}\boldsymbol{\tilde{D}}_{1}^{\dagger} - \boldsymbol{I}                                            & \boldsymbol{\tilde{D}}_{1}^{\dagger} \left( \boldsymbol{\tilde{D}}_{3'}^{\dagger} \boldsymbol{\tilde{D}}_{3}^{\dagger} - \boldsymbol{I}\right) & \boldsymbol{\tilde{D}}_{2'}^{\dagger} -\boldsymbol{\tilde{D}}_{1}^{\dagger}\boldsymbol{\tilde{D}}_{3}^{\dagger} \\
				\boldsymbol{0}                                                                                             & \boldsymbol{0}                                                                                                & \boldsymbol{I} - \boldsymbol{\tilde{D}}_{1}^{\dagger}\boldsymbol{\tilde{D}}_{2}^{\dagger}\boldsymbol{\tilde{D}}_{3}^{\dagger} \\
				\boldsymbol{0}                                                                                             & \boldsymbol{I} - \boldsymbol{\tilde{D}}_{1}^{\dagger}\boldsymbol{\tilde{D}}_{2}^{\dagger}\boldsymbol{\tilde{D}}_{3}^{\dagger}                        & \boldsymbol{0} \\
				\boldsymbol{I} - \boldsymbol{\tilde{D}}_{1}^{\dagger}\boldsymbol{\tilde{D}}_{2}^{\dagger}\boldsymbol{\tilde{D}}_{3}^{\dagger}                    & \boldsymbol{0}                                                                                               & \boldsymbol{0}
			\end{pmatrix}. 
		\end{equation}
	\end{widetext}
	Note that $\boldsymbol{\tilde{V}}_{n}$ is also a basis for the null space of the laser-noise part of the covariance, so that we have $\boldsymbol{\tilde{\Sigma}} \boldsymbol{\tilde{V}}_{n} = \boldsymbol{S}_{n} \boldsymbol{\tilde{V}}_{n}.$
	
\begin{figure}[ht]
		\centering
		\includegraphics[width=0.49\textwidth, trim={0.5cm, 0.5cm, 0.1cm, 0.4cm}, clip]{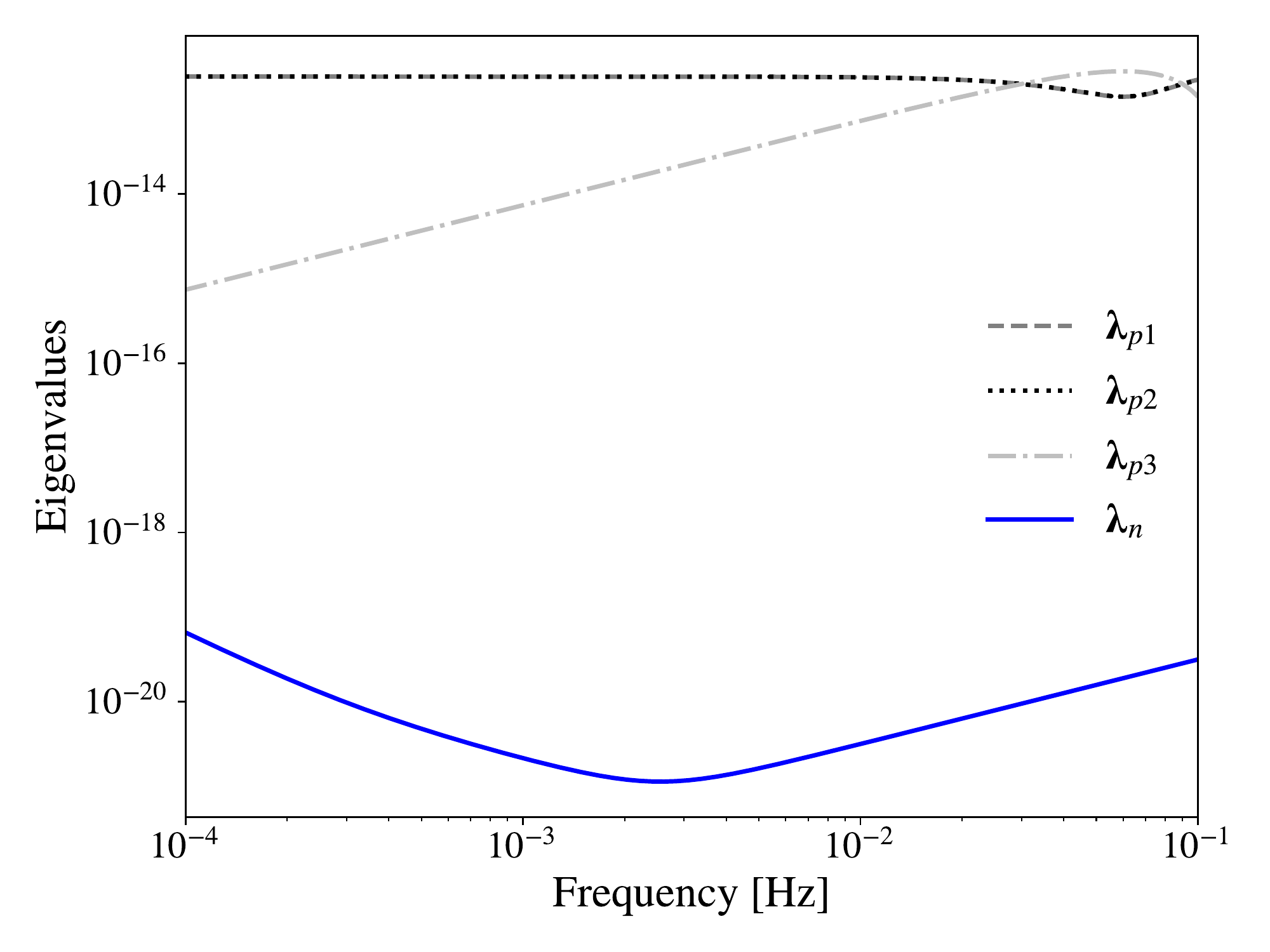}
		\caption{Laser-noise dominated eigenvalues $\boldsymbol{\lambda}_{pi}$ (gray) and laser-noise free eigenvalues $\boldsymbol{\lambda}_{n}$ (blue) of the phasemeter covariance matrix as a function of frequency.}
		\label{fig:laser_cov_eigenvalues}
	\end{figure}

	The $3 N_{f}$ other eigenvalues, that we label as $\boldsymbol{\Lambda}_{p} = \mathrm{diag}\left(\boldsymbol{\lambda}_{p1} , \boldsymbol{\lambda}_{p2}, \boldsymbol{\lambda}_{p3}\right)$, have more complicated expressions but are all proportional to the laser noise PSD $S_{p}$. We denote by $\boldsymbol{\tilde{V}}_{p}$ the associated eigenvector matrix, which has the same dimensions as $\boldsymbol{\tilde{V}}_{n}$. We plot the laser-noise dominated eigenvalues $\boldsymbol{\lambda}_{pi}$ in gray as a function of frequency in Fig.~\ref{fig:laser_cov_eigenvalues}, along with the degenerate laser-noise free eigenvalues $\boldsymbol{\lambda}_{n}$ in blue. This figure confirms that the former are much larger than the latter.

	As a result, we can partition the eigenvector matrix as $\boldsymbol{\tilde{V}} = \begin{pmatrix} \boldsymbol{\tilde{V}}_{p} & \boldsymbol{\tilde{V}}_{n} \end{pmatrix}$. Let us consider the data transformation $\boldsymbol{\tilde{e}} \equiv \boldsymbol{\tilde{V}}^{\dagger} \boldsymbol{\tilde{y}}$. As they correspond to different eigenvalues, the eigenvector matrices $\boldsymbol{\tilde{V}}_{n}$ and $\boldsymbol{\tilde{V}}_{p}$ are orthogonal. Therefore, the covariance of $\boldsymbol{\tilde{e}}$ is block diagonal:
	\begin{eqnarray}
		\label{eq:cov_e}
		\mathrm{Cov}\left( \boldsymbol{\tilde{e}} \right) = \begin{pmatrix}
			\boldsymbol{V}_{p}^{\dagger} \boldsymbol{\Sigma} \boldsymbol{V}_{p} & \boldsymbol{0} \\ 
			\boldsymbol{0} & \boldsymbol{V}_{n}^{\dagger} \boldsymbol{S}_{n} \boldsymbol{V}_{n}
		\end{pmatrix}
		\equiv 
		\begin{pmatrix}
			\boldsymbol{C}_{p} & \boldsymbol{0} \\ 
			\boldsymbol{0} & \boldsymbol{C}_{n}
		\end{pmatrix},
	\end{eqnarray}
	where we defined the $3 N_{f} \times 3 N_{f}$ matrix $\boldsymbol{C}_{n}$ (respectively $\boldsymbol{C}_{p}$) as the covariance of the projected data $\boldsymbol{\tilde{e}}_{n} \equiv \boldsymbol{\tilde{V}}_{n}^{\dagger} \boldsymbol{\tilde{y}}$ (respectively $\boldsymbol{\tilde{e}}_{p} \equiv \boldsymbol{\tilde{V}}_{p}^{\dagger} \boldsymbol{\tilde{y}}$) onto the laser-noise free (respectively laser-noise dominated) basis.
	
	Then it is possible to separate the laser noise-dominated eigenbasis $\boldsymbol{\tilde{V}}_{p}$ from the laser noise-free eigenbasis $\boldsymbol{\tilde{V}}_{n}$ in the calculation of the likelihood:
	\begin{eqnarray}
		\label{eq:likelihood_separated}
		\log p\left( \boldsymbol{y}  |  \boldsymbol{\theta} \right) & = & - \frac{1}{2} \left(\boldsymbol{\tilde{y}} -  \boldsymbol{\tilde{h}} \right)^{\dagger} \boldsymbol{\tilde{V}}_{p} \boldsymbol{C}^{-1}_{p} \boldsymbol{\tilde{V}}^{\dagger}_{p} \left(\boldsymbol{\tilde{y}} -  \boldsymbol{\tilde{h}} \right)
		\nonumber \\
		&& - \frac{1}{2} \left(\boldsymbol{\tilde{y}} -  \boldsymbol{\tilde{h}} \right)^{\dagger} \boldsymbol{\tilde{V}}_{n} \boldsymbol{C}^{-1}_{n} \boldsymbol{\tilde{V}}^{\dagger}_{n} \left(\boldsymbol{\tilde{y}} -  \boldsymbol{\tilde{h}} \right)
		\nonumber \\
		&&  - \frac{1}{2} \left( \log \left| \boldsymbol{C}_{p} \right| + \log \left| \boldsymbol{C}_{n} \right| \right).
	\end{eqnarray}
	
	Similarly to Romano and Woan's example in Ref.~\cite{Romano2006}, the laser noise variance being much larger than other noises, the term in the second line in Eq.~(\ref{eq:likelihood_separated}) is almost constant as a function of $\boldsymbol{\theta}$ compared to the term in the first line. Therefore, \hl{for parameter inference purposes}, we can safely approximate the log-likelihood by
	\begin{eqnarray}
		\label{eq:likelihood_approx}
		\log p\left( \boldsymbol{y}  |  \boldsymbol{\theta} \right) & \approx & - \frac{1}{2} \left(\boldsymbol{\tilde{y}} -  \boldsymbol{\tilde{h}} \right)^{\dagger} \boldsymbol{\tilde{V}}_{n} \boldsymbol{C}^{-1}_{n} \boldsymbol{\tilde{V}}^{\dagger}_{n} \left(\boldsymbol{\tilde{y}} -  \boldsymbol{\tilde{h}} \right)
		\nonumber \\
		&&  - \frac{1}{2} \log \left| \boldsymbol{C}_{n} \right|.
	\end{eqnarray}
	In the next section, we detail how we compute the inverse of $\boldsymbol{C}_{n}$.

	\subsection{\label{sec:pci_non_laser_noises}Orthogonalization with respect to non-laser noise}
	
	In the previous section, we saw that the approximate log-likelihood depends on the covariance $\boldsymbol{C}$ of the data projected onto the eigenbasis associated with null eigenvalues of the laser-noise covariance. If the noise parameters (PSD levels and light travel time delays) are assumed to be known and fixed in the inference scheme, this matrix, and its inverse, can be computed once for all. However, if we need to update the delays and the noise model along the way, we must compute $\boldsymbol{C}^{-1}_{n}$ at every parameter update for all frequencies. In this work, we perform this computation by numerically diagonalizing $\boldsymbol{C}_{n}$ with its eigenvectors $\boldsymbol{\Phi}$ and eigenvalues $\boldsymbol{\Lambda}_{C}$, and then computing $\boldsymbol{C}^{-1}_{n} = \boldsymbol{\Phi} \boldsymbol{\Lambda}_{C}^{-1} \boldsymbol{\Phi}^{\dagger}$. We use the $\textsc{NumPy}$ library~\citep{TravisE2006, VanDerWalt2011}, which includes efficient algorithms when the number of frequency bins is not large ($< 1000$). It may be more efficient for larger frequency bands to use an analytical formula for $3 \times 3$ Hermitian matrices as derived by~Ref. \cite{Kopp2006}.
	
	This diagonalization is a generalization of the orthogonalization process that leads to TDI variables A, E, T~\citep{Estabrook2000, Tinto2014}. Indeed, we can apply the formalism developed in this section to any linear transformation that cancels laser frequency noise. For example, TDI transformations can be encoded by some matrix $\boldsymbol{T}$ instead of $\boldsymbol{V}_{n}$ in Sect.~\ref{sec:laser_noise_only}. 
	While channels A, E, and T are orthogonal under specific conditions (including equal armlengths, non-rotating constellation, identical acceleration noise levels, and uncorrelated noises), the rationale leading to Eq.~(\ref{eq:likelihood_approx}) does not rely on any of these assumptions.

	\subsection{\label{sec:unknown_noises}PCI with no prior knowledge of noise PSDs}
	
	While the projection onto the null space of the laser noise covariance matrix (i.e., the calculation of $\boldsymbol{\tilde{V}_{n}}$) is independent of the laser noise spectra, the orthogonalization that we outlined in Sect.~\ref{sec:pci_non_laser_noises} relies on our knowledge of the other noises' spectral matrix $\boldsymbol{S}_{n}$. Although we may have a physical model describing acceleration and OMS noises, we must expect deviations from the theory when dealing with future LISA data. Therefore, it is necessary to have a formalism that also allows us to estimate $\boldsymbol{C}_{n}$ robustly. Estimating the full covariance matrix elements is not commonly done in gravitational-wave data analysis, but it can be performed in similar to PSD estimation methods, extending them to off-diagonal terms. This type of problem relates to spectral analysis of co-stationary multivariate time series, for which several approaches are available, such as estimating the components of the generalized Cholesky decomposition of the spectral matrix or its inverse~\citep{Dai2004}. Regardless of the model we adopt, one has to ensure that the estimated spectrum is a positive definite matrix and is continuous as a function of frequency. To benefit from fast conditional steps when it comes to posterior sampling, we choose to model the covariance elements themselves with the regression scheme proposed by Ref.~\cite{Hoff2012}, which takes advantage of conjugate priors. To this end, let us consider a single frequency $f$, and the $3 \times 3$ covariance of the corresponding elements:
	\begin{equation}
		\label{eq:covariance_vs_frequency}
		\boldsymbol{\check{C}}_{n}(f) \equiv \mathrm{Cov}\left(\boldsymbol{\check{e}}(f)\right),
	\end{equation}
	where we labeled as $\boldsymbol{\check{e}}(f_{k}) \equiv \left(\tilde{y}_{k},\, \tilde{y}_{N_{f}+k},\, \tilde{y}_{2N_{f}+k}\right)^{T}$ the vector of eigenstream elements associated with frequency bin $f_{k}$.
	Thus, we assume that the covariance has the form
	\begin{eqnarray}
		\label{eq:covariance_model}
		\boldsymbol{\check{C}}_{n}(f) = \boldsymbol{\Psi} + \boldsymbol{B} \boldsymbol{x}(f) \boldsymbol{x}^{\dagger}(f) \boldsymbol{B}^{\dagger},
	\end{eqnarray}
	where $\boldsymbol{\Psi}$ is a constant $3 \times 3 $ Hermitian matrix, $\boldsymbol{x}(f)$ is a $q \times 1$ design matrix depending on frequency, and $\boldsymbol{B}$ is a $3 \times q $ matrix of regression parameters. For example, $\boldsymbol{x}$ can have the form of a polynomial in frequency with elements $\boldsymbol{x} = \left(1,\, f, \, \hdots, \,   f^{q-1} \right)^{T}$.
	
	In this model, $\boldsymbol{\Psi}$ and $\boldsymbol{B}$ are unknown and must be estimated. For a sufficiently short frequency range, we can even approximate the covariance by a constant term across the band, as in Ref.~\cite{Littenberg2020}. Under this assumption, Eq.~(\ref{eq:covariance_model}) reduces to $\boldsymbol{\check{C}}_{n}(f) = \boldsymbol{\Psi}$, and Ref.~\cite{Hoff2012}'s sampling scheme amounts to using the conjugate prior for the Gaussian distribution, i.e. the inverse-Wishart prior $\mathrm{IW}\left(\boldsymbol{\Psi}_{0}, \nu_{0}\right)$. We adopt this simplification in what follows, where the conditional posterior of $\boldsymbol{\Psi}$ given the delays $\boldsymbol{L}$ and GW parameters $\boldsymbol{\theta}_{\mathrm{GW}}$ is also inverse-Wishart: 
	\begin{eqnarray}
		p\left(\boldsymbol{\Psi}| \boldsymbol{y}, \boldsymbol{L}, \boldsymbol{\theta}_{\mathrm{GW}}\right) = \mathrm{IW}\left( \boldsymbol{\Psi}_{0} + \hat{\boldsymbol{\Psi}}, \nu_{0} + N_{f} \right),
	\end{eqnarray}
	where $N_{f}$ is the number of frequency bins and $\hat{\boldsymbol{\Psi}}$ is the $3 \times 3$ sample covariance of the eigenstream residuals:
	\begin{eqnarray}
		\hat{\boldsymbol{\Psi}} = \boldsymbol{\tilde{V}}_{n}^{\dagger} \left(\boldsymbol{\tilde{y}} - \boldsymbol{\tilde{h}}\right)   \left(\boldsymbol{\tilde{y}} - \boldsymbol{\tilde{h}}\right)^{\dagger} \boldsymbol{\tilde{V}}_{n}.
	\end{eqnarray}
	We implement this step in \textsc{Python} using statistical packages from the \textsc{Scipy} library~\citep{Virtanen2020}. Following Ref.~\cite{Hoff2012}'s suggestions we set $\nu_{0} = d + 2$, where $d = 3$ is the dimension of $\boldsymbol{\Psi}_{0}$. We choose $\boldsymbol{\Psi}_{0}$ to be the median of the frequency bins' sample covariances after a first run obtaining a rough estimate of the eigenstreams $\boldsymbol{\tilde{e}}_{n}$.

	\subsection{\label{sec:frequency_implementation}Frequency-domain implementation of delays}
	
	Up to now, we assumed that time series have a quasi-infinite length so that the asymptotic frequency-domain formulation of the delay operator in Eq.~(\ref{eq:asymptotic_delay}) is valid. In practice, we analyze relatively short measurements for which this approximation breaks. Applying Eq.~(\ref{eq:asymptotic_delay}) on Fourier-transformed data leads to large edge effects. To mitigate this behavior, we use a time-window that smoothly drops to zero at the time series' edges. However, such an operation usually requires transforming the data back to the time domain, which is computationally expensive compared to the usual cost of one likelihood evaluation. Therefore, we perform the equivalent computation in Fourier space (i.e., a discrete convolution) using a sparse approximation of the convolution kernel, similarly as in covariance approximation techniques~\citep{Furrer2006}. The delay operation amounts to the following matrix multiplication:
	\begin{eqnarray}
		\boldsymbol{\tilde{D}}_{\mathrm{tap}} = \boldsymbol{\tilde{\Omega}} \boldsymbol{\tilde{D}},
	\end{eqnarray}
	where $\boldsymbol{\tilde{D}}$ is the asymptotic delay operator as given by Eq.~(\ref{eq:covariance_matrix_fourier}) and $\boldsymbol{\tilde{\Omega}}$ is the tapered convolution matrix whose elements are given by
	\begin{equation}
		\tilde{\Omega}_{k, p}  = \left\{
		\begin{array}{ll}
			\sum_{n=0}^{N-1} w(n) e^{-2 \pi j n\frac{k - p}{N}}  & \mbox{if } | k - p | \leq p_{0} ;\\
			0 & \mbox{otherwise.}
		\end{array}
		\right.
	\end{equation}
	We denoted by $w(n)$ the time-domain window function and $p_{0}$ an integer threshold for the row-column difference, above which the matrix elements are zero.

	\section{\label{sec:simulations}Case study}
	
	To demonstrate the developed approach's performance, we consider the simple case where phasemeter data only contain a single GW source buried into stationary Gaussian noise. Unless otherwise stated, we assume arbitrary armlengths and noise PSDs. The parameters governing the estimation model are:
	\begin{itemize}
		\item Laser light travel time delays \\
		$\boldsymbol{L} = \left( L_{1} ,\, L_{2} ,\,  L_{3} ,\,  L_{1'} ,\, L_{2'} ,\,  L_{3'} \right)$;
		\item GW source parameters that we restrict to intrinsic ones $\boldsymbol{\theta}_{\mathrm{GW}} = \left( \theta,\, \phi,\,  f_{0} ,\,  \dot{f}_{0} \right)$;
		\item Non-laser noise covariance parameters $\boldsymbol{\Psi}$.
	\end{itemize}
	We use the likelihood function in Eq.~(\ref{eq:likelihood_approx}) that we maximize over extrinsic GW amplitudes. In this function, the frequency-domain waveform $\boldsymbol{\tilde{h}}$ depends both on $\boldsymbol{\theta}_{\mathrm{GW}}$ and on $\boldsymbol{L}$, while the laser noise covariance eigenvectors $\boldsymbol{V}_{n}$ depend on $\boldsymbol{L}$ only.

	\subsection{\label{sec:parametrization}Simulation parametrization}
	
	\paragraph{Noise.}
	We simulate one month of LISA observations that yield single-link time series by implementing Eq.~(\ref{eq:phasemeneter_meas}) with a \textsc{Python} code. We generate noises at a sampling cadence of 2 Hz. We first applied the delays using time-domain Lagrange interpolation filters, with the same parametrization as in LISANode~\citep{Bayle2019}, and checked that the PCI algorithm was successfully canceling laser noise. 
	However, we noted that the applied delays' accuracy was not enough to be unnoticed when recovering delays from data simulated over long periods (one month). In other words, the delay values optimally canceling laser noise were slightly biased compared to injected delay values. This mismatch is understandable, as fractional delay filters have a frequency response that only approximates the ideal delay filter~\citep{Tseng2010}. Therefore, we chose to simulate the data used in this study directly in the frequency domain, relying on noise stationarity, following~Ref.~\cite{Timmer1995}.
	
	We then filter the data using a Kaiser finite-response filter and downsample it to 0.2 Hz to generate the outputs. We assume a rigid, rotating LISA constellation so that the effective armlengths do not vary in time and that the light travel time is sensitive to the direction of propagation due to the Sagnac effect.
	
	The noise PSD model includes three components: laser frequency noise, test-mass (TM) acceleration noise, and optical metrology system (OMS) noise. We assume that the noises affecting two different optical benches are uncorrelated, so that matrices $\boldsymbol{S}_{p}$ and $\boldsymbol{S}_{n}$ are block diagonal. Matrix $\boldsymbol{S}_{n}$ has 6 diagonal blocks $\boldsymbol{S}_{n_{i}}$ of the form:
	\begin{eqnarray}
		{S}_{n_{i}| k, l} = \alpha_{i} \left( S_{\mathrm{TM}}(f_{k}) + S_{\mathrm{OMS}}(f_{k}) \right) \delta_{kl},
	\end{eqnarray}
	where $\alpha_{i}$ is a positive coefficient depending on optical bench $i$. Expressions for noise PSDs $S_{\mathrm{TM}}(f)$ and $S_{\mathrm{OMS}}(f)$ are given in Appendix~\ref{sec:psds}. Thus, noise spectra have the same shape for every optical benches, up to a coefficient accounting for possible noise level discrepancies.
	
	\paragraph{Gravitational-wave signal.}
	We assume that the gravitational signal comes from the loudest verification compact galactic binary known to date, called HM Cnc~\citep{Stroeer2006, Kupfer2018}. We simulate single-link gravitational-wave signals sampled at 0.2 Hz in the time domain using the same code as in Ref.~\citep{Baghi2019a} that we adapted by removing the TDI transfer function. We also relaxed the low-frequency approximation, using a Fourier series decomposition similar to Cornish and 
	Littenberg's implementation in Ref.~\cite{Cornish2007}. For all parameters, we use uniform priors around the true parameter values. The source's characteristics, along with prior boundaries, are summarized in Table~\ref{tab:source_parameters}.
	\begin{table}[ht]
		\caption{\label{tab:source_parameters}Values of the source parameters used in the simulations, with their uniform prior boundaries.}
		\begin{ruledtabular}
			\begin{tabular}{l c c}
				Parameter & Value & Prior range\\ \hline
				\\[-0.75em]
				Frequency [mHz] & 6.22 & $\left[6.12,\, 6.32\right]$ \\
				Frequency derivative [$\rm mHz / s$] & \hl{7.49 $\times 10^{-13}$} & \hl{$\left[10^{-14},\, 10^{-10}\right]$}\\
				Ecliptic latitude [rad] & -0.0821 & \hl{$\left[-\pi/2,\, \pi/2\right]$} \\
				Ecliptic longitude [rad] & 2.102 & $\left[-\pi,\, \pi\right]$ \\
				Amplitude [strain] &  $\rm 6.4 \times 10^{-23}$ & Marginalized\\
				Inclination [rad] & 0.6632 & Marginalized \\
				Initial phase [rad] & 5.78 & Marginalized\\
				Polarization [rad] & 3.97 & Marginalized\\
			\end{tabular}
		\end{ruledtabular}
	\end{table}
	
	\subsection{Projection onto the null space}
	The first simulation we consider includes single-link measurements where all noises are generated from the same PSD, as given in Appendix~\ref{sec:psds}. It also contains one single GW source, as described in Sect.~\ref{sec:parametrization}. For the sake of description, here we assume that the light travel time delays $L_{i} / c$ are known. We compute the null-space eigenvector matrix $\boldsymbol{\tilde{V}}_{n}$ analytically using Eq.~(\ref{eq:constant_delay_eigenvectors}) and the other eigenvector matrix $\boldsymbol{\tilde{V}}_{p}$ numerically. Thanks to these matrices, we apply the PCI transformation to obtain eigenstreams that we orthogonalize as described in Sect.~\ref{sec:pci_non_laser_noises}. We label as $\boldsymbol{\tilde{e}}_{\perp i}$ these orthogonal streams.
	\begin{figure}[ht]
		\centering
		\includegraphics[width=0.48\textwidth, trim={0.5cm, 0.4cm, 0.4cm, 0.5cm}, clip]{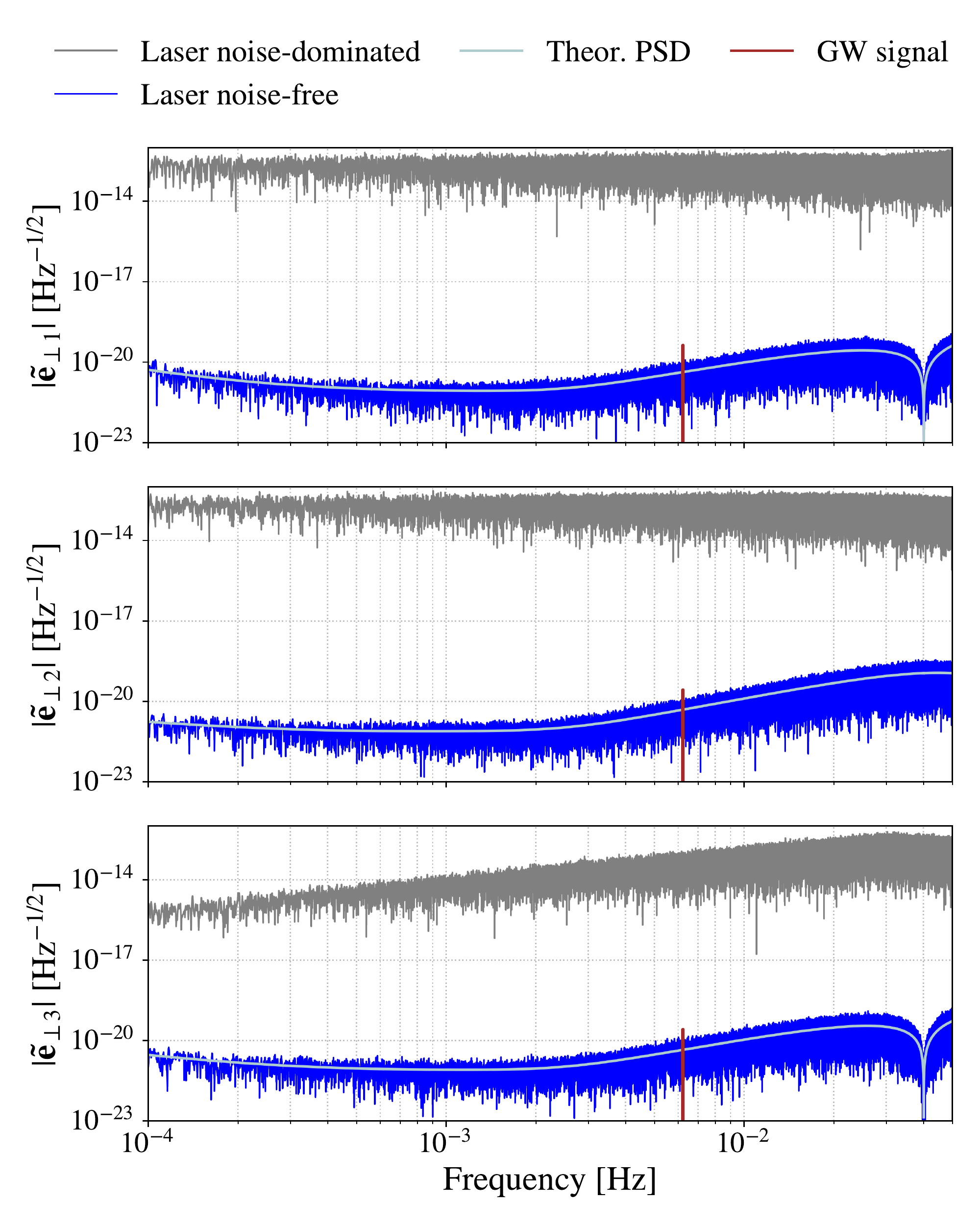}
		\caption{PCI eigenstreams expressed in relative frequency deviation for the laser-noise free subspace (blue) and the laser-noise dominated subspace (gray) for a month-long simulation. The theoretical PSD function of laser-noise free eigenstreams are plotted in light blue. The red vertical lines denote the gravitational signal from the verification galactic binary HM Cnc that shows up in all channels as a quasi-monochromatic signal.}
		\label{fig:pci_equal_noises}
	\end{figure}
	
	We plot the periodogram of the PCI transformations in Fig.~\ref{fig:pci_equal_noises}, which shows the laser-noise dominated eigenstreams (gray) along with the null space eigenstreams (blue), lying 8 orders of magnitude below. 
	We also plot the GW signal transformed in the null space eigenstreams, which emerges from the noise. We show that this noise is indeed limited by the acceleration and OMS errors by plotting the theoretical PSD (light blue) computed from the diagonal elements of the covariance matrix $\mathrm{Cov}\left(\boldsymbol{\tilde{e}}_{\perp}\right) = \boldsymbol{\Phi}^{\dagger} \boldsymbol{\tilde{V}}_{n}^{\dagger} \boldsymbol{S}_{n}\boldsymbol{\tilde{V}}_{n} \boldsymbol{\Phi}$. These plots demonstrate the ability of the frequency-domain algorithm to separate the two orthogonal spaces correctly.

	\subsection{\label{sec:sensitivity_analysis}Sensitivity analysis}
	
\begin{figure}[ht]
		\includegraphics[width=0.49\textwidth, trim={0.5cm, 0.5cm, 0.1cm, 0.4cm}, clip]{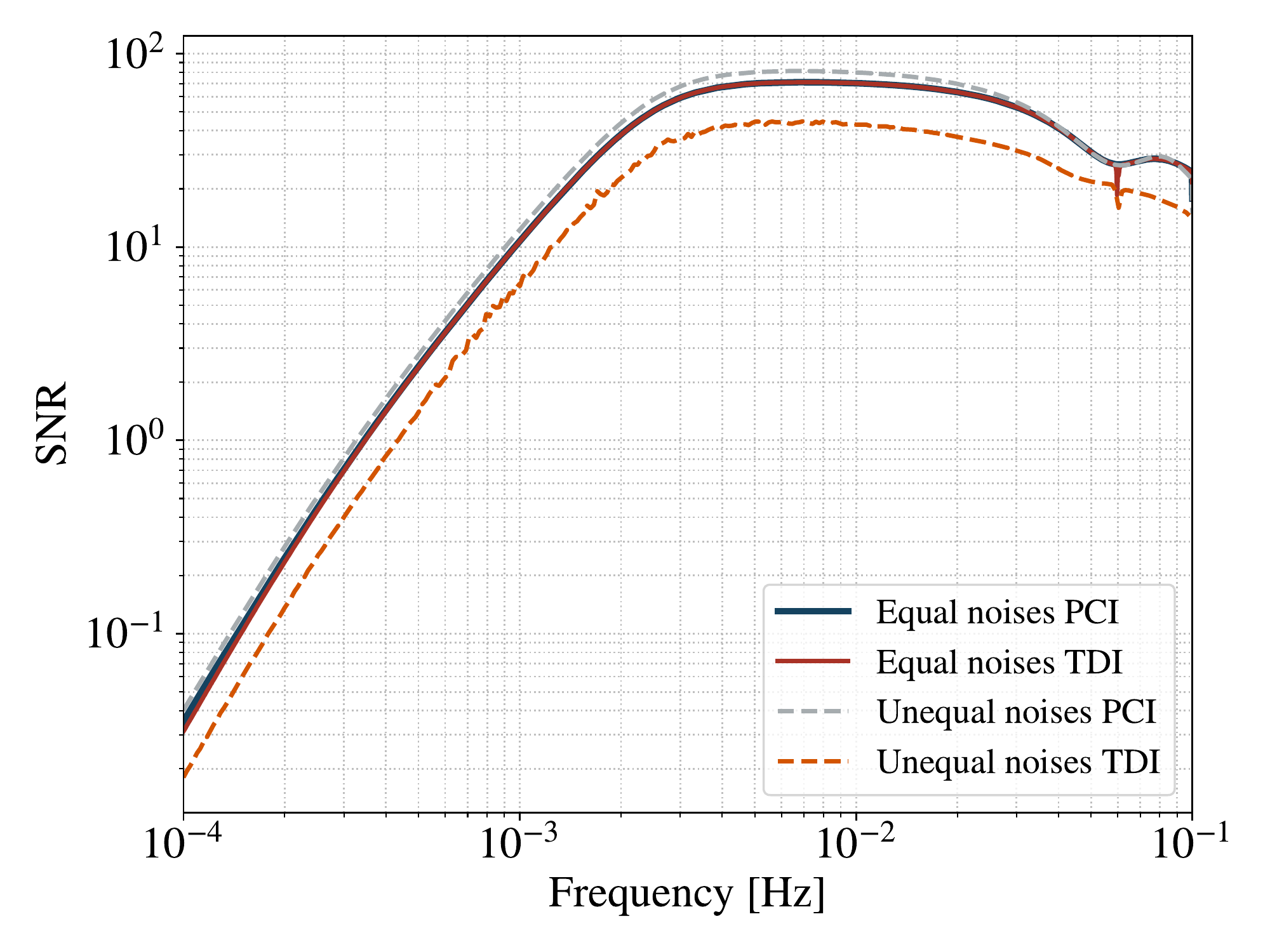}
		\caption{ \label{fig:sensitivities} Sensitivity of laser-noise free channels as a function of frequency obtained with PCI eigenstreams (blue or gray) and unoptimized TDI channels A, E, T (red or orange). The solid  lines represent the case of equal noise; dashed lines correspond to unequal noise levels, computed for a source like HM Cnc. PCI and TDI sensitivities are almost identical in the case of equal noises, while unoptimized TDI undergoes a significant SNR loss due to misorthogonalization in the case of unequal noises.}
	\end{figure}

	In this section we investigate the theoretical performance of the PCI approach in two cases: i) all non-laser noise levels are the same, i.e., $\alpha_{i} = 1\, \forall i$ and ii) non-laser noise levels are different depending on optical benches, with $\bm{\alpha} = \left(4,\, 0.25,\,  16,\, 0.1,\, 0.4,\, 1 \right)$.
	In case ii), some optical bench noises have larger amplitudes than the baseline, while others have smaller amplitudes. Overall, the mean noise level is larger than in case i) by a factor 3.6. In Fig.~\ref{fig:sensitivities}, we plot in blue the generalized sensitivity to an ultra-compact galactic binary source \hl{observed over one year}. \hl{The source we consider has the same location as HM Cnc, with zero inclination and equal polarization modes; and we allow its frequency to vary. Here, for any frequency $f$, ``sensitivity" refers to the signal-to-noise ratio (SNR) of a source of frequency $f$ measured in its full bandwidth.} This calculation takes into account the fact that the covariance matrix is non-diagonal in general, as described in Appendix~\ref{sec:sensitivity_computation}. 
	For comparison, we plot in red and orange generalized sensitivities of unoptimized TDI combinations A, E, and T. They are obtained from combining TDI Michelson X, Y, and Z strictly as derived in~\citep{Prince}, relying on the assumption of equal noises and equal arm lengths. Thus, we compute associated sensitivities assuming that A, E, T's covariance matrix is perfectly diagonal; hence we call them ``unoptimized".  
	
	PCI and unoptimized TDI yield almost the same sensitivity when noises are equal, as shown by the superimposition of the blue and red, solid curves. This similarity confirms that the A, E, T formulation is nearly optimal in this configuration because the assumptions made in their derivation are almost met, except for the equal arm lengths hypothesis, which plays a minor part in the orthogonalization process.
	
	\hl{The figure shows that the unequal-noise case (dashed lines) yields a different SNR compared to the case where all noises are equal (solid lines). However, the change in SNR is smaller with the PCI method as it includes the frequency-dependent orthogonalization by design, providing a general extension to classic TDI. For this particular source, the value yielded by the PCI decomposition with unequal noises is even slightly larger than with equal noises. This is because some channels have a larger noise level while others have a smaller one. This comparison illustrates the importance of off-diagonal terms in the covariance $\boldsymbol{C}_{n}$, which are not equal in the case of heterogeneous noise levels. Thus, not taking this discrepancy into account may result in sub-optimal performance.}
	
	\subsection{\label{sec:inference_scheme}Parameter inference scheme}
	
	We further assess the performance of the developed method by using it to recover injected parameters from the numerical simulations described in Sect.~\ref{sec:parametrization}. Estimated parameters include light travel time delays, source parameters, and noise covariance. 
	We restrict the inference to a portion of the frequency band \hl{between $6.20$ and $6.24$ mHz} around the binary's frequency. 
	We sample the posterior distribution of delays and GW parameters through parallel-tempered Markov Chain Monte Carlo (PTMCMC) sampling, using the \textsc{ptemcee} algorithm~\citep{Vousden2015}, a parallel-tempered version of the affine-invariant ensemble sampler \textsc{emcee}~\citep{Foreman-Mackey2012}. 
	
	We modify the sampling algorithm to include noise covariance parameters in the inference, using a two-step Blocked Gibbs sampling scheme where noise parameters are sampled conditionally to delays and GW parameters:
	\begin{eqnarray}
		\label{eq:sampling_scheme}
		\text{Step 1: } && \boldsymbol{L}, \boldsymbol{\theta}_{\mathrm{GW}} \sim p\left(\boldsymbol{L}, \boldsymbol{\theta}_{\mathrm{GW}}  | \boldsymbol{y}, \boldsymbol{\Psi}, \boldsymbol{B}\right) ; \nonumber \\ 
		\text{Step 2: } && \boldsymbol{\Psi} \sim  p\left(\boldsymbol{\Psi}| \boldsymbol{y}, \boldsymbol{L}, \boldsymbol{\theta}_{\mathrm{GW}}\right).
	\end{eqnarray}
	While step 1 is still based on PTMCMC, Step 2 uses direct sampling as described in Sect.~\ref{sec:unknown_noises}. We describe sampling results in the next section.

	\subsection{\label{sec:inference}Inference results}
	
	We present the results of the inference of delays, GW parameters, and covariance parameters applied to the synthetic data described in Sect.~\ref{sec:parametrization} with the sampling scheme presented in Sect.~\ref{sec:inference_scheme}. \hl{To obtain sufficient precision, we extend the simulation duration to one year, sampled at 20~mHz.}
	
	First, we use two data sets: one corresponding to the equal noise case i) described in Sect.~\ref{sec:sensitivity_analysis}, the other for the unequal noise case ii). We run 40 chains in parallel with 10 different temperatures, and we retain $4 \times 10^{5}$ samples after chains have reached convergence. 
	
\begin{figure}[ht]
		\centering
		\includegraphics[width=0.47\textwidth, trim={0.5cm, 0cm, 0cm, 0cm}, clip]{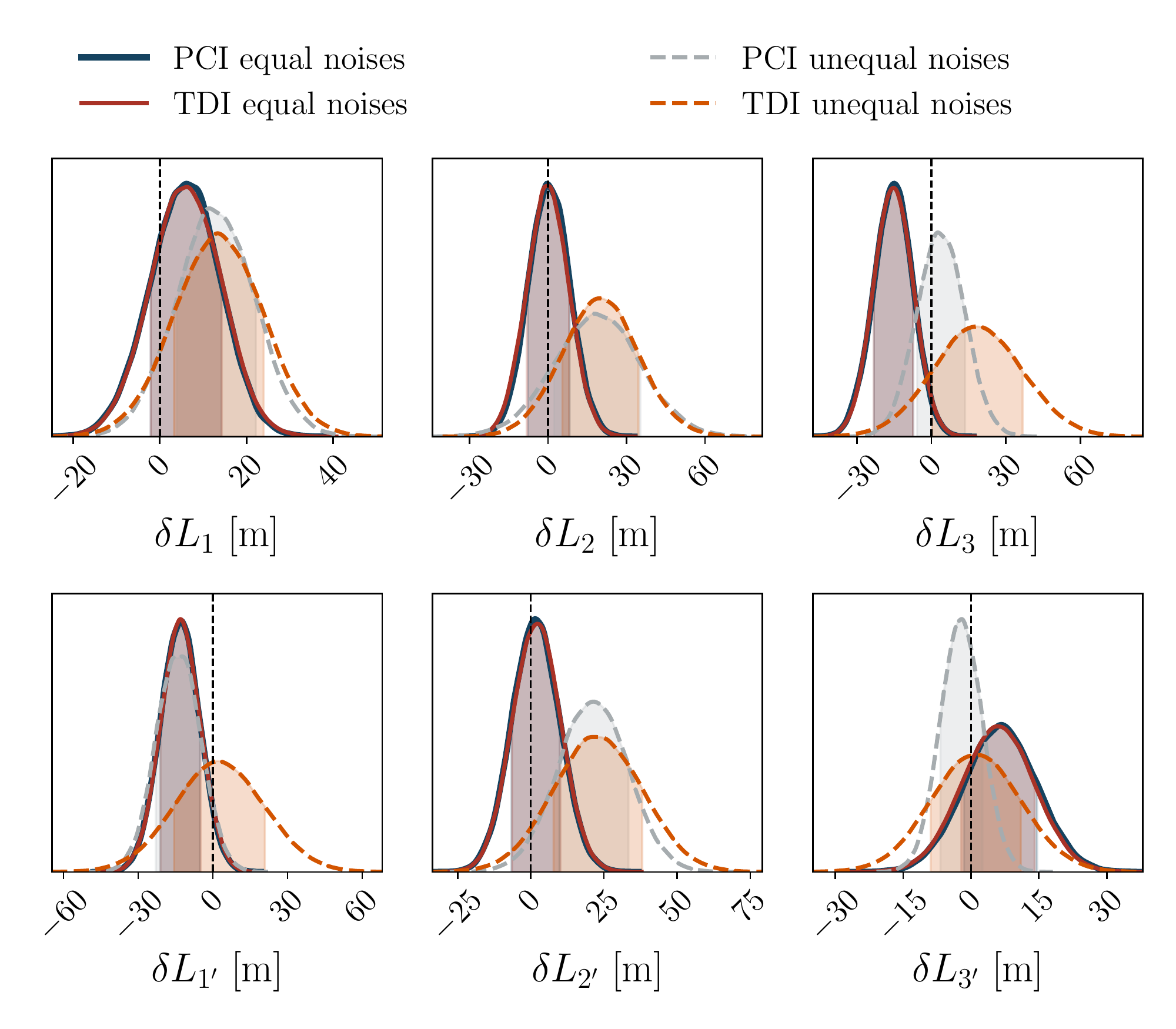}
		\caption{Posterior distribution of the 6 light-travel time delays expressed in equivalent arm lengths. Solid lines correspond to the case of equal acceleration noises, whereas dashed lines correspond to unequal noises. Posteriors obtained with PCI are in blue or gray, and posteriors obtained with unoptimized TDI are in red or orange. Thin vertical black dashed lines represent true values, and shaded areas under the curves cover the $1\sigma$ region.}
		\label{fig:delay_posterior}
	\end{figure}

	\paragraph{Delays.} We plot in Fig.~\ref{fig:delay_posterior} the delays posteriors marginalized over other parameters in the case of equal (dashed lines) and unequal (solid lines) acceleration noises, using the \textsc{ChainConsumer} package~\citep{Hinton2016}. We express delays in equivalent inter-spacecraft distances. In the case of equal noise levels, posteriors obtained from PCI and unoptimized TDI are almost equal to each other, confirming the result found in Fig.~\ref{fig:sensitivities}. 
	
	Delays distributions are broader in the case of unequal noises because, in this example, the overall noise power is larger than in the case of equal noises. However, for most delays, posteriors have a larger variance with classic TDI than with the PCI analysis, because PCI accounts for the change of off-diagonal covariance terms, maintaining orthogonalization. This result is consistent with the significant SNR loss shown in Fig.~(\ref{fig:sensitivities}) when using unoptimized TDI combinations.
	
	We remark that the uncertainty in estimating equivalent arm lengths is of order 40 m, which is enough to cancel laser noise in this particular case. Should we wish to, we could obtain a better precision in using the entire frequency data instead of restricting it to a narrow band.

\begin{figure}[ht]
		\centering
		\includegraphics[width=0.47\textwidth, trim={0cm, 0cm, 0cm, 0cm}, clip]{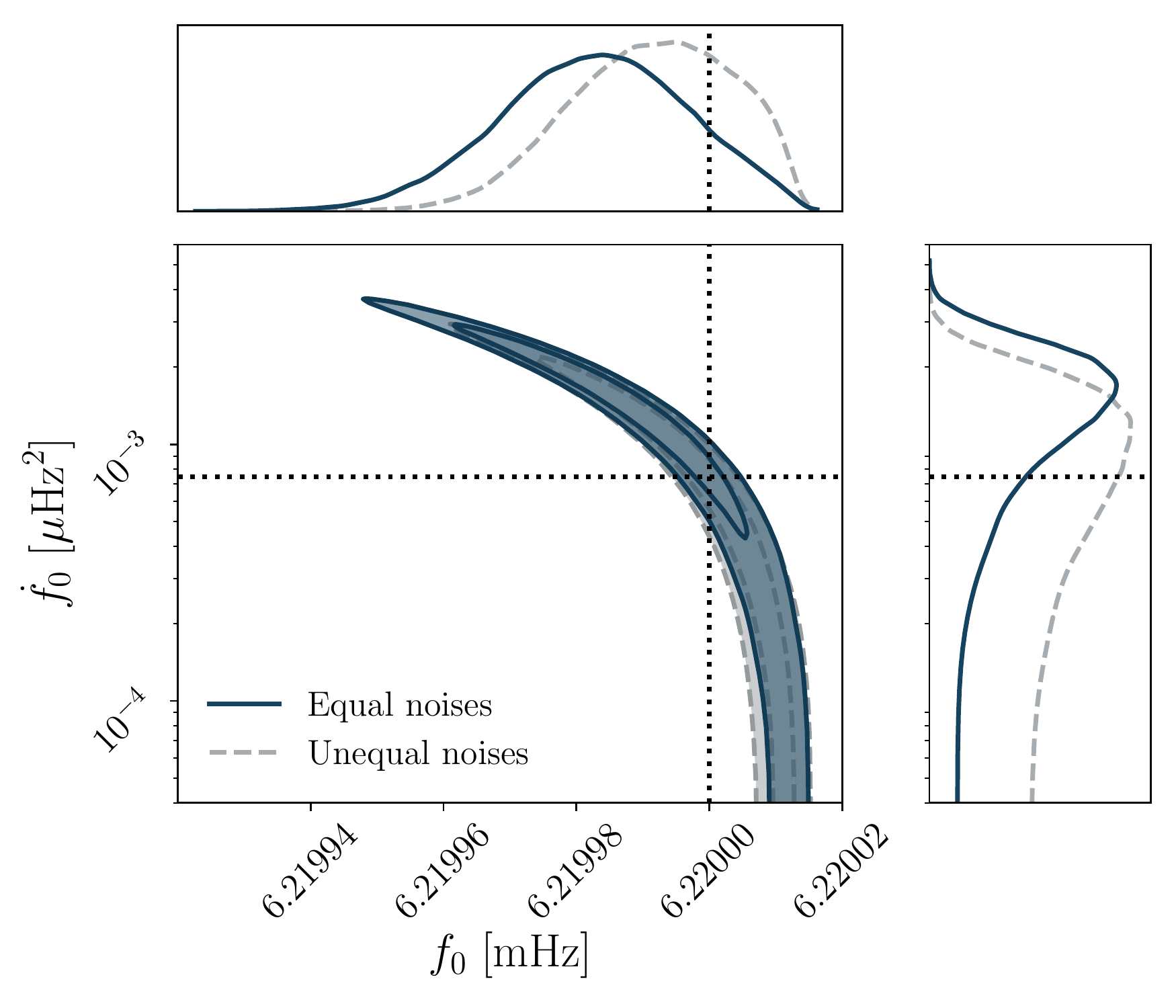}
		\caption{Joint posterior distribution of GW source's frequency $f_{0}$ and frequency derivative $\dot{f}_{0}$, obtained with PCI in the case of equal (solid blue) and unequal (dashed gray) noises applied to a one-month long simulation of phasemeter measurements. Contours correspond to $1\sigma$ and $2\sigma$ regions.}
		\label{fig:frequency_posterior_pci}
	\end{figure}

	\paragraph{GW parameters.}
	Then, in Fig.~\ref{fig:frequency_posterior_pci}, we examine the posterior of the GW source's frequency and frequency derivative, marginalized over all other parameters. Here we focus on PCI results only, comparing equal (solid blue lines) and unequal (gray dashed lines) noises. 
	\hl{The figure shows that the frequency is accurately recovered by the PCI analysis (within about 1 $\rm \mu Hz$), even when performed simultaneously with the estimation of laser light delays. We observe a minor difference between equal and unequal noise cases, showing that the adaptive orthogonalization built in the PCI process minimizes the impact of the noise heterogeneity.}
	
\begin{figure}[ht]
		\centering
		\includegraphics[width=0.49\textwidth, trim={0.6cm, 0.5cm, 0.3cm, 0.5cm}, clip]{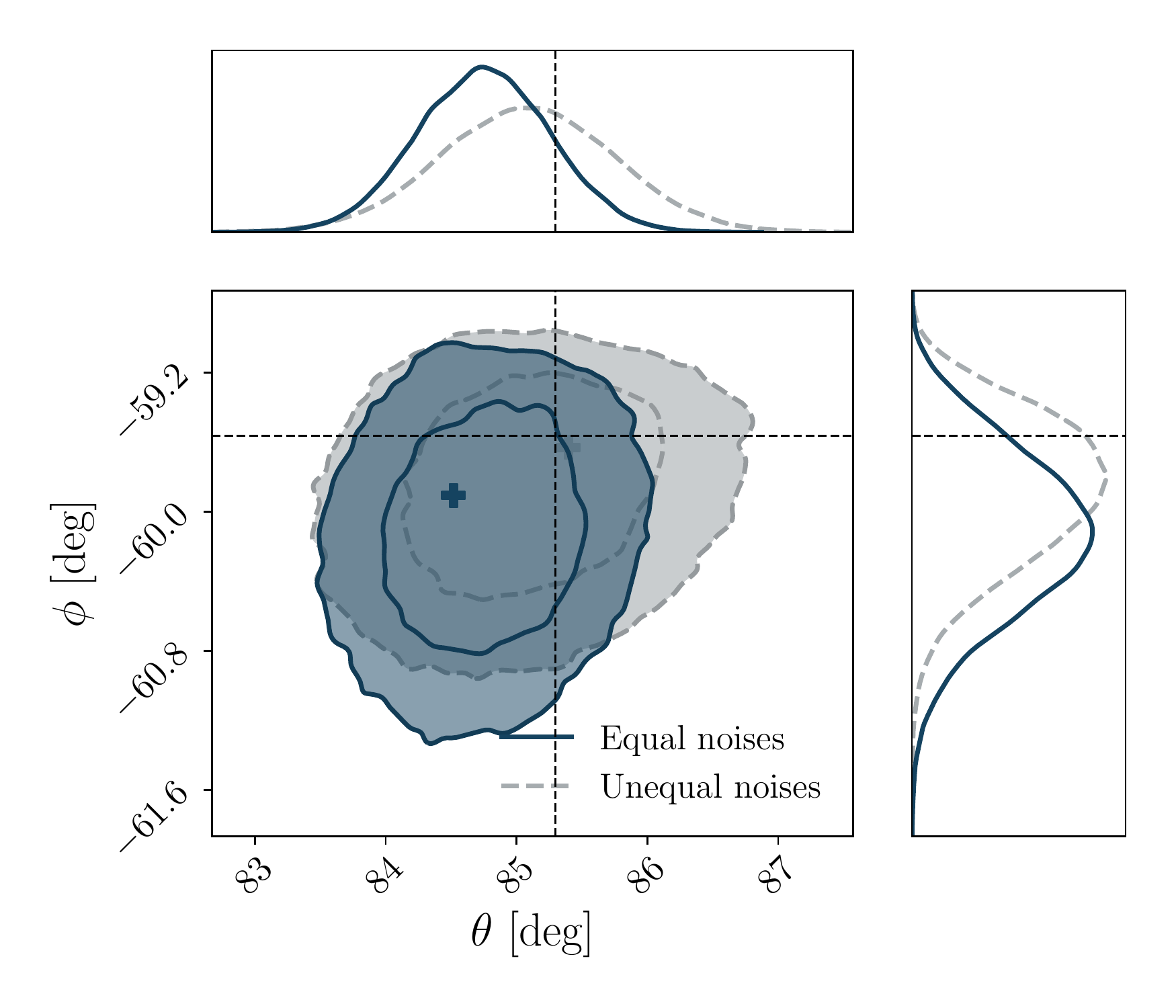}
		\caption{Sky localization posterior distribution of the GW source in celestial coordinates, obtained with PCI in the case of equal (solid blue) and unequal (dashed gray) noises. The dashed lines represent the true location.} 
		\label{fig:sky_posterior_pci}
	\end{figure}

	 \hl{The GW source sky location posteriors that we plot in Fig}.~\ref{fig:sky_posterior_pci} \hl{exhibit the same behavior as for the frequency parameters. The joint PCI analysis accurately spots the source's location in the sky (dashed black lines), as shown by the maximum a posteriori estimate (MAP) represented by the blue (equal noises) and gray (unequal noises) crosses. Hence, the MAP estimate remains close to the actual sky location even in the case of unequal noises. }

\begin{figure}[ht]
		\centering
		\includegraphics[width=0.47\textwidth, trim={0.5cm, 0cm, 0cm, 0cm}, clip]{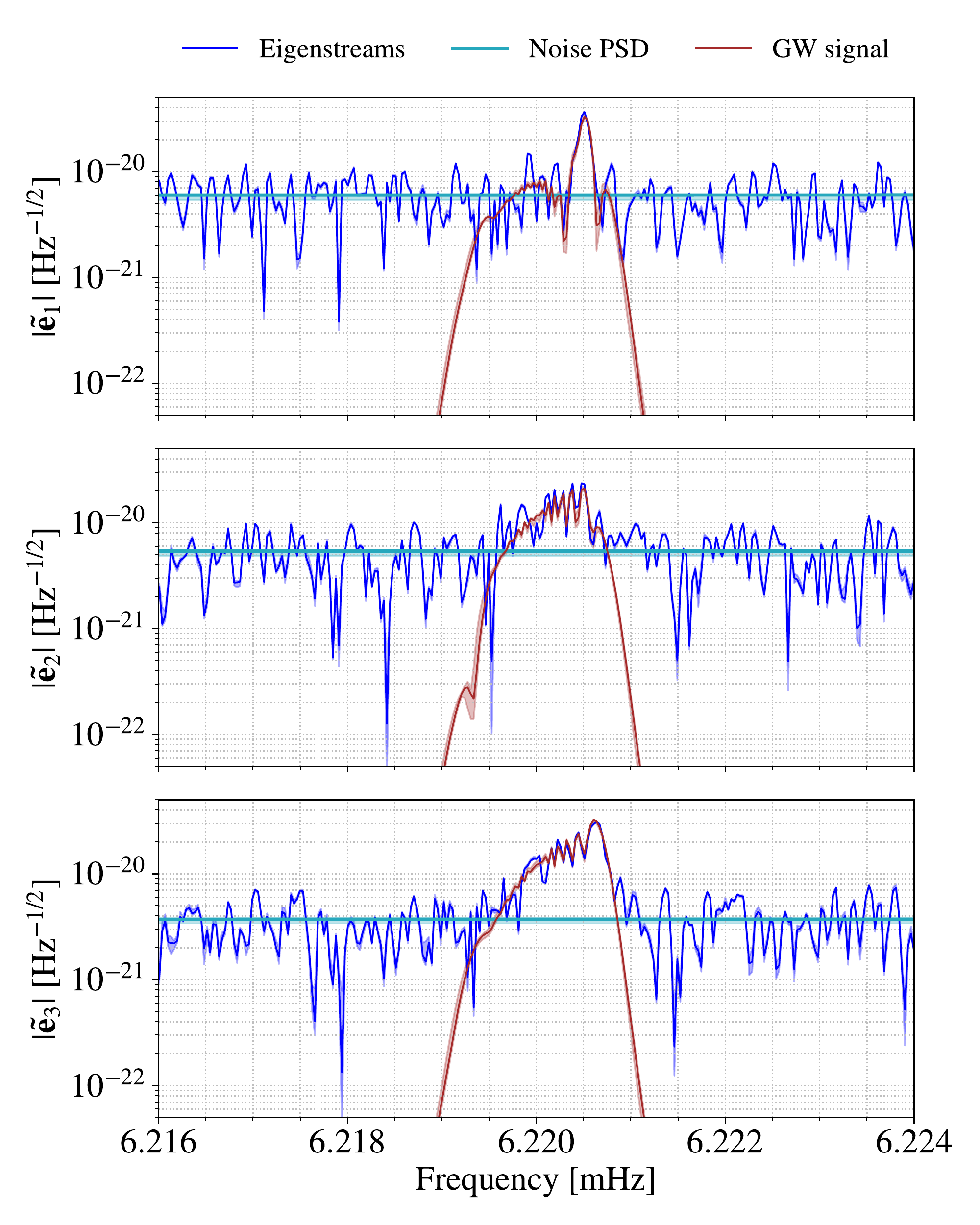}
		\caption{True value (solid lines) and estimated $3\sigma$ posterior interval (light shaded areas) of the eigenstream periodograms (blue), the GW signal (red) and the noise PSDs (diagonal covariance entries) expressed in relative frequency deviation. Posterior intervals are computed using 1000 MCMC samples. The data shown here corresponds to the case of unequal non-laser noises.}
		\label{fig:covariance_posterior}
	\end{figure}

	\paragraph{Covariance.} We collect the covariance parameter samples computed in the Gibbs steps in Fig.~\ref{fig:covariance_posterior}. We use them to compute the chains of covariance values. We plot the estimated posterior of the diagonal terms in light green in Fig.~\ref{fig:covariance_posterior} as a $3\sigma$ interval around the mean, and we compare it to the true value represented by the solid line.
	The real value is located within the $3\sigma$ interval, demonstrating the covariance estimate's accuracy. We obtain similar-looking plots for off-diagonal covariance elements (real and imaginary parts), confirming the accurate characterization of non-laser noise and the proper orthogonalization of the eigenstreams. 
	
	We also plot the periodograms of the three eisgenstreams using the delays' actual values (solid blue line), along with the $3\sigma$ interval of the posterior samples. The posterior closely encompasses the target value, showing that the ranging estimates' variability is acceptable.
	
	Finally, we plot the $3\sigma$ interval (red shaded area) of the GW waveform samples against the true value of the signal (solid red curve), showing the right consistency between the two. As a result, Fig.~\ref{fig:covariance_posterior} provides a summary of the multi-parameter inference enabled by the PCI framework.
	
	\section{\label{sec:discussion}Discussion and prospects}

	We revisited space-based GW data modeling by writing the model likelihood directly as a function of phasemeter measurements. Based on Romano and Woan's idea in Ref.~\cite{Romano2006}, we restricted the likelihood to its lowest variance terms using principal component analysis. This work's main contribution is to formalize the PCA approach in the frequency domain, using an asymptotic formulation of the delay operators with sparse matrices. This formalization provides a framework that is readily applicable to parameter inference, yielding optimal precision. We show that it allows us to handle the TDI transfer function of both signal and noise in a single compact, matrix-based formalism. 
	
	With a simplified example of simulated LISA data, we show that the method allows us to consistently and simultaneously fit for inter-spacecraft phase delays, an ultra-compact galactic binary source's parameters, and noise covariance parameters. We show that a numerical and data-driven diagonalization of the covariance yields an optimal sensitivity to gravitational waves and minimal source parameter uncertainty. 
	
	This work lays the foundation for a more robust analysis of LISA data. First, it provides a way to derive sensitivities from instrumental noises, tracking all correlations straightforwardly. Second, it generalizes the concept of orthogonal TDI variables to arbitrary armlengths and noise correlations. However, to apply to in-orbit data, the approach must be extended to time-varying inter-spacecraft distances. Future work will focus on formalizing this time-dependence in the frequency domain to maintain computational efficiency. Furthermore, we can extend the developed framework to a complete set of LISA measurements by dropping the assumption of identical spacecraft's laser noise and include reference interferometer measurements. Preliminary work has already demonstrated the successful decomposition into large and low variance components in this configuration, that we will present in follow-up studies.

	Finally, our work highlights the importance of L0 data including raw phasemeter measurements in LISA's science analysis.
	The availability of such data will ensure that alternative processing methods complementary to the standard TDI pipeline are possible. This diversity of approaches is an essential tool for cross-checking and validating the data in an off-line interferometry step that is crucial for the precise characterization of astrophysical sources.

	\appendix
	\section{\label{sec:psds}Expressions of noise PSD model}
	
	The noise PSD models used in this study are given in fractional frequency per Hertz ($\mathrm{Hz}^{-1}$) as
	\begin{eqnarray}
		S_{\mathrm{laser}}(f) & = & \left(\frac{a_0}{\nu_0}\right)^{2} \;;\nonumber \\
		S_{\mathrm{TM}}(f) & = &  \left(\frac{a_{\mathrm{TM}}}{ 2 \pi f c}\right)^2 
		\left(1 + \left(\frac{f_{1}}{f}\right)^2\right) \left(1 + \left(\frac{f}{f_{2}}\right)^4\right) \;; \nonumber \\
		S_{\mathrm{OMS}}(f) & = & a_{\mathrm{OMS}}^{2} \left( \frac{2 \pi f }{c} \right)^2 \left(1 + \left(\frac{f_{3}}{f}\right)^4 \right) .
	\end{eqnarray}
	We indicate the values of the noise model parameters in Table~\ref{tab:noise_parameters}.
	\begin{table}[ht]
		\caption{\label{tab:noise_parameters}Values of the noise parameters used in the simulations.} 
		\begin{ruledtabular}
			\begin{tabular}{l l l}
				Noise type & Parameter & Value \\ \hline
				\\[-0.8em] 
				\multirow{2}{*}{Laser} &  $a_0$ & 28.2 $\rm Hz.{Hz}^{-1/2}$ \\ 
				& $\nu_0$ & 281759 $\rm GHz$ \\ \hline 
				\\[-0.8em]
				\multirow{3}{*}{Test-mass} & $a_{\mathrm{TM}} $ & 3 $\mathrm{fm s^{-2} {Hz}^{-1/2}}$  \\
				& $f_{1}$ &  0.4 mHz \\
				& $f_{2}$ & 8 mHz \\ \hline
				\\[-0.8em]
				\multirow{2}{*}{OMS} & $a_{\mathrm{OMS}}$ & 15 $\mathrm{pm{Hz}^{-1/2}}$  \\
				&  $f_{3}$ & 2 mHz
			\end{tabular}
		\end{ruledtabular}
	\end{table}
	
	\section{\label{sec:sensitivity_computation}Computation of sensitivity to monochromatic binaries}
	
	In this section, we derive the expression for the generalized SNR plotted in Fig.~\ref{fig:sensitivities}, which extends the classic SNR calculation to the case where the noise covariance used for the estimation is different than the correct one.
	
	Let us consider some linear transformation of the single-link measurements, encoded by a \hl{$6 N_f \times 3 N_f$} matrix $\boldsymbol{W}$ that can represent any TDI or PCI transformation in the frequency domain, \hl{over a bandwidth of $N_f$ frequency bins}. 
	Applying $\boldsymbol{W}$ to the phasemeter measurements \hl{vector $\boldsymbol{\tilde{y}}$ of size $6N_f$ yields the vector $\boldsymbol{\tilde{e}}$ of size~$3N_f$}. 
	
	We define the generalized SNR $\rho_{\boldsymbol{W}}(f_{0})$ of a monochromatic GW of frequency $f_{0}$ obtained with the data transformation $\boldsymbol{W}$ as the ratio between the absolute value of its gravitational-wave strain amplitude $h_{0}$ and the standard deviation $\sigma_{0}$ of its maximum likelihood estimate:
	\begin{eqnarray}
		\rho_{\boldsymbol{W}}(f_{0}) \equiv \frac{|h_{0}|}{\sigma(f_0)}.
	\end{eqnarray}
	
	We assume that the GW signal appears in the single-link measurements $\boldsymbol{\tilde{y}}$ as 
	\begin{eqnarray}
		\boldsymbol{\tilde{y}} = \boldsymbol{\tilde{A}} h_{0},
	\end{eqnarray}
	where $\boldsymbol{\tilde{A}}$ is a \hl{$6 N_f \times 1$} design matrix.
	From Eq.~(\ref{eq:observation_equation}), this transformation maps the GW amplitude as 
	\begin{eqnarray}
		\boldsymbol{\tilde{e}} = \boldsymbol{W}^{\dagger} \boldsymbol{\tilde{A}} h_{0} + \boldsymbol{\tilde{\epsilon}},
	\end{eqnarray}
	where $\boldsymbol{\tilde{\epsilon}}$ is the noise contribution to the data stream vector $\boldsymbol{\tilde{e}}$. The effective $3 N_f \times 1$ transfer function matrix is thus $\boldsymbol{H} \equiv \boldsymbol{W}^{\dagger} \boldsymbol{\tilde{A}}$.
	
	Let us assume that we analyse the data with the following likelihood model:
	\begin{equation}
		\label{eq:likelihood_model_h0}
		p\left( \boldsymbol{\tilde{e}} |  h_{0} \right) = \frac{\exp\left\{- \frac{1}{2} \left(\boldsymbol{\tilde{e}}- \boldsymbol{H} h_{0} \right)^{\dagger}\boldsymbol{C}^{-1}\left(\boldsymbol{\tilde{e}}- \boldsymbol{H} h_{0} \right)\right\}}{\sqrt{(2 \pi)^{6N}  \left| \boldsymbol{C} \right| }},
	\end{equation}
	where $\boldsymbol{C}$ is some model of the covariance of $\boldsymbol{\tilde{e}}$.
	Assuming that $\boldsymbol{C}$ is fixed, the maximum likelihood estimator error on $h_{0}$ from data \hl{in the bandwidth of size $N_f$} is then given by the inverse Fisher matrix, which is defined as
	\begin{equation}
		\label{eq:inverse_fisher}
		\sigma^{2}_{0} = \mathrm{E} \left[\left( \frac{\partial \log p}{\partial h_{0}} \right)^{2} \right]^{-1}.
	\end{equation}
	Inserting Eq.~(\ref{eq:likelihood_model_h0}) into Eq.~(\ref{eq:inverse_fisher}) yields
	\begin{eqnarray}
		\sigma^{2}_{0} = \frac{\boldsymbol{H}^{\dagger} \boldsymbol{C}^{-1} \boldsymbol{\Sigma}_{e} \boldsymbol{C}^{-1} \boldsymbol{H}}{\left(\boldsymbol{H}^{\dagger} \boldsymbol{C}^{-1} \boldsymbol{H}\right)^{2}},
	\end{eqnarray}
	where $\boldsymbol{\Sigma}_{e}$ is the true variance of data streams $\boldsymbol{\tilde{e}}$, and is given by
	\begin{eqnarray}
		\boldsymbol{\Sigma}_{e} = \boldsymbol{W}^{\dagger}\boldsymbol{\tilde{\Sigma}} \boldsymbol{W},
	\end{eqnarray}
	where $\boldsymbol{\tilde{\Sigma}}$ is the covariance of $\boldsymbol{\tilde{y}}$.
	We can remark 2 properties:
	\begin{itemize}
		\item If $\boldsymbol{W}$ is an orthogonal transformation with respect to the noise, then $\boldsymbol{\Sigma}_{e}$ is diagonal by construction.
		\item If $\boldsymbol{C} = \boldsymbol{\tilde{\Sigma}}_{e}$ (i.e., the model covariance is equal to the true one), then $\rho_{\boldsymbol{W}}(f_{0})  =|h_{0}| \sqrt{\boldsymbol{H}^{\dagger} \boldsymbol{\Sigma}_{e}^{-1} \boldsymbol{H}}$ which is exactly equivalent to the standard SNR formula used in the gravitational-wave literature.
	\end{itemize}
	
	\begin{acknowledgments}
		We would like to thank Tyson Littenberg and Jessica Page for enlightening conversations. We would also like to acknowledge interesting feedback from Martin Staab and Shane Larson. This work is supported by an appointment to the NASA Postdoctoral Program at the Goddard Space Flight Center, administered by Universities Space Research Association under contract with NASA.
	\end{acknowledgments}
	
	\bibliography{myreferences}
	
\end{document}